\UseRawInputEncoding
   
\documentclass{article}
\usepackage{eurosym}
\usepackage{amssymb}
\usepackage{amsmath}
\usepackage{cite}
\usepackage{graphicx}
\usepackage{epsf}
\usepackage{lscape}
\usepackage{epstopdf}
\usepackage{fancyhdr}

\begin{document}

\title{Group properties and solutions for the 1D Hall MHD system in the cold
plasma approximation}
\author{Andronikos Paliathanasis \\
{\ \textit{Institute of Systems Science, Durban University of Technology }}\\
{\ \textit{PO Box 1334, Durban 4000, Republic of South Africa}} \\
{\textit{Instituto de Ciencias F\'{\i}sicas y Matem\'{a}ticas,}}\\
{\ \textit{Universidad Austral de Chile, Valdivia, Chile}}\\
Email: anpaliat@phys.uoa.gr}
\maketitle

\begin{abstract}
We study the Lie point symmetries and the similarity transformations for the
partial differential equations of the nonlinear one-dimensional
magnetohydrodynamic system with the Hall term known as HMHD system. For this
1+1 system of partial differential equations we find that is invariant under
the action of a seventh dimensional Lie algebra. Furthermore, the
one-dimensional optimal system is derived while the Lie invariants are
applied for the derivation of similarity transformations. We present
different kinds of oscillating solutions.\newline

\end{abstract}

\section{Introduction}

\label{sec1}

Lie point symmetries play a significant role in the study on nonlinear
differential equations. The novelty of Lie point symmetries is that they
provide invariant functions which define similarity transformations such
that to write the differential equation in an equivalent simplest form. The
latter can be done in two-ways, either by reducing the number of independent
variables, in the case of partial differential equations, or by reduce the
order of the differential equation, in the case of ordinary differential
equations. Furthermore, Lie symmetries can be used to classify differential
equations with common group properties and when it is feasible to find an
equivalent transformation between different kinds of differential equations
invariants under the elements of same Lie algebra.

In the theory of fluid dynamics Lie symmetries has been widely studied and
applied in various problems, and they have been used for the determination
and the classification of these hydrodynamic systems. Applications of the
Lie symmetries on the Shallow-water equations with or without the Coriolis
force have been studied before in \cite%
{swr06,sw1,sw2,sw3,sw4,sw5,sw6,sw7,mel1,mel2,mel3,mel4}. The group
properties of the two-phase flows system were studied in \cite{tw9,tw10,tw11}%
. In the presence of an electromagnetic field, and specifically in Plasma
physics Lie symmetries has been applied before. The group properties for the
ideal equations of magnetohydrodynamic (MHD) were studied in detailed in
\cite{mh1}. The Lie point symmetries for the MHD convection flow and heat
transfer of an incompressible viscous nanofluid past a semi-infinite
vertical stretching sheet in the presence of thermal stratification were
studied in \cite{mh2}.\ The Grad--Shafranov equation which is and
equilibrium equation in ideal MHD was studied on the existence of point
symmetries in \cite{mh3}. The relation of the Lie point symmetries and
Noether symmetries and the Lagrangian map in MHD were investigated before in
\cite{mh4,mh5,mh6,mh7}. On the other hand, new solutions were found recently
for the Pulsar equation in \cite{mh8}.

In this work we are interested on the study of the group-properties of the
MHD equations where the Hall term effect is included. In ideal MHD the Hall
term is very small and usually is neglected, thus Hall term play a
significant role in the study of the magnetic reconnection due to its
ability to accurately describe plasmas with large magnetic field gradients
\cite{mh13}, while other important applications of the Hall term in MHD can
be found for instance in \cite{mh9,mh10,mh11,mh12}.

The Hall MHD (HMHD) equations without any pressure term of the ions or any
electron pressure are \cite{mh13}%
\begin{eqnarray}
\rho _{,t}+\nabla \left( \rho \mathbf{u}\right)  &\mathbf{=}&0\mathbf{,}
\label{hm.01} \\
\left( \rho \mathbf{u}\right) _{,t}+\nabla \left( \rho \mathbf{u~u}+\frac{%
\left\vert B\right\vert ^{2}}{2}-\mathbf{BB}\right)  &=&0,  \label{hm.02} \\
\mathbf{B}_{,t}-\nabla \times \left( -u\times B+\frac{\xi _{0}}{\rho }\left(
\mathbf{\nabla \times B}\right) \mathbf{\times B}\right)  &=&0.
\label{hm.03}
\end{eqnarray}%
where $\xi _{0}$ is the coefficient parameter for the Hall term, while when $%
\xi _{0}=0$ the equations of MHD are recovered. The zero pressure
consideration is also known as cold plasma approximation. The dimensionless
parameter $\xi _{0}$ is the ion inertial scale or skin depth defined as $\xi
_{0}=\frac{c}{\omega _{i}L}$, where $c$ is the speed of light, $\omega _{i}$
is the ion plasma frequency and $L$ is the scale length of the plasma.\ The
Alf\'{e}n speed is defined as $V_{A}=\frac{B_{0}}{\sqrt{\rho _{0}\mu }}$
where for the vacuum permeability we have assumed $\mu =1$. If someone
considered parallel propagation on the magnetic field the HMHD system for
large values of the Hall term $\xi _{0}$ the Alfv\'{e}n waves propagates in
the fast manifold by the derivative nonlinear Schr\"{o}dinger equation
(DNLS) \cite{nd01,nd02,nd03}. As it was found before in \cite{nd1} the DNLS
is an integrable equation while its algebraic properties has been studied
before in \cite{nd2}. Specifically in \cite{nd2} the triple degenerate
nonlinear Schr\"{o}dinger system (TDNLS) was studied. TDNLS arises from wave
propagation along the magnetic field, when the gas sound speed matches the
Alfv\'{e}n speed, that is, the slow and Alfv\'{e}n speeds coincide. Because
the original DNLS equation has a singular, divergent coefficient for the
nonlinear term at this limit, a modified perturbation approach has been
considered in order the difference between the Alfv\'{e}n speed and sound
speed to be assumed small in the perturbation analysis.

In a highly ionized plasma the Hall effect follows because of the difference
in electron and ion inertia. Specifically, ions are incapable to follow the
magnetic fluctuations at frequencies higher than their cyclotron, while
electrons remain coupled to the magnetic field lines. An important
characteristic of the HMHD system is the it admits a Hamiltonian
formulation. In \cite{hhal} the authors defined a set of canonical variables
to describe an equivalent canonical Hamiltonian system with the HMHD system,
while this property was used to recover the MHD limit. An alternative
approach on the construction of the noncanonical Poisson brackets for the
HMHD system can be found in \cite{hhal1}.

We continue by considering that the system is one dimensional and the
magnetic field is constant on the direction $x$, thus let us assume~$\rho
=\rho \left( t,x\right) $, $u=u^{1}\left( t,x\right) \partial
_{x}+u^{2}\left( t,x\right) \partial _{y}+u^{3}\left( t,x\right) \partial
_{z}~$and $B=B_{0}^{1}\partial _{x}+B^{2}\left( t,x\right) \partial
_{y}+B^{3}\left( t,x\right) \partial _{z}$ such that the HMHD equations are
simplified in the following form \cite{vvs}%
\begin{eqnarray}
\rho _{,t}+\left( \rho u^{1}\right) _{,x} &=&0,  \label{hm.04} \\
\left( \rho u^{1}\right) _{,t}+\left( \rho \left( u^{1}\right) ^{2}+\frac{1}{%
2}\left( \left( B^{2}\right) ^{2}+\left( B^{3}\right) ^{2}\right) \right)
_{,x} &=&0,  \label{hm.05} \\
\left( \rho u^{2}\right) _{,t}+\left( \rho u^{1}u^{2}-B^{1}B^{2}\right)
_{,x} &=&0,  \label{hm.06} \\
\left( \rho u^{3}\right) _{,t}+\left( \rho u^{1}u^{3}-B^{1}B^{3}\right)
_{,x} &=&0,  \label{hm.07} \\
B_{,t}^{2}-\left( -\left( u^{1}B^{2}-u^{2}B_{0}^{1}\right) +\xi _{0}\frac{%
B_{0}^{1}}{\rho }B_{,x}^{3}\right) _{,x} &=&0,  \label{hm.08} \\
B_{,t}^{3}+\left( -\left( u^{3}B_{0}^{1}-u^{1}B^{3}\right) +\xi _{0}\frac{%
B_{0}^{1}}{\rho }B_{,x}^{2}\right) _{,x} &=&0.  \label{hm.09}
\end{eqnarray}%
where for the $x$-component from the Faraday's equaiton it follows $%
B_{0,t}^{1}=0$ and $B_{0,x}^{1}=0$. For the latter system we study the
admitted Lie point symmetries, the invariants of the admitted Lie algebra
are investigated as also the invariants are applied for the derivation of
similarity solutions. The plan of the paper is as follows.

In Section \ref{sec2} we apply Lie's theory and we derive the infinitesimal
generators, i.e. the Lie symmetries, for the one-parameter point
transformation which leaves\ the system (\ref{hm.04})-(\ref{hm.09})
invariant. Specifically, we found that the system admits seven Lie point
symmetries, one symmetry less from the same system in the ideal MHD without
the Hall-term. The commutators and the Adjoint representation for the
admitted Lie symmetries are determined which are used to determine the
one-dimensional optimal system. In Section \ref{sec4} we demonstrate the use
of the Lie symmetries by presenting the application of some similarity
transformations which lead to integrable\ reduced systems. We recover
previous results for the existence of solitary waves as also we find new
oscillating solutions. Finally, in Section \ref{sec5} we summarize our
results and we draw our conclusions.

\section{Lie symmetries for the 1D HMHD equations}

\label{sec2}

Consider the infinitesimal one-parameter point transformation%
\begin{eqnarray}
\bar{t} &=&t+\varepsilon \xi ^{1}\left( t,x,\rho ,\mathbf{u}%
,B^{2},B^{3}\right) ,~\bar{x}=x+\varepsilon \xi ^{2}\left( t,x,\rho ,\mathbf{%
u},B^{2},B^{3}\right) ,  \label{hm.10} \\
\bar{\rho} &=&\rho +\varepsilon \eta ^{1}\left( t,x,\rho ,\mathbf{u}%
,B^{2},B^{3}\right) ,~\bar{u}^{1}=u^{1}+\varepsilon \eta ^{2}\left( t,x,\rho
,\mathbf{u},B^{2},B^{3}\right) ,  \label{hm.12} \\
\bar{u}^{2} &=&u^{2}+\varepsilon \eta ^{3}\left( t,x,\rho ,\mathbf{u}%
,B^{2},B^{3}\right) ,~\bar{u}^{2}=u+\varepsilon \eta ^{4}\left( t,x,\rho ,%
\mathbf{u},B^{2},B^{3}\right) ,  \label{hm.13} \\
B^{2} &=&B^{2}+\varepsilon \eta ^{5}\left( t,x,\rho ,\mathbf{u}%
,B^{2},B^{3}\right) ,B^{3}=B^{3}+\varepsilon \eta ^{6}\left( t,x,\rho ,%
\mathbf{u},B^{2},B^{3}\right) .  \label{hm.14}
\end{eqnarray}%
with $\mathbf{X}=\xi ^{1}\partial _{t}+\xi ^{2}\partial _{x}+\eta
^{1}\partial _{\rho }+\eta ^{2}\partial _{u^{1}}+\eta ^{3}\partial
_{u^{2}}+\eta ^{4}\partial _{u^{3}}+\eta ^{5}\partial _{B^{2}}+\eta
^{6}\partial _{B^{3}}$. Hence, we shall say that the HMHD system $\mathbf{%
H\equiv 0}$ defined by the equations (\ref{hm.04})-(\ref{hm.09}) is
invariant under the Action of the latter one-parameter point transformation
if and only if%
\begin{equation}
\mathbf{X}^{\left[ 1\right] }\left( \mathbf{H}\right) =0,  \label{hm.15}
\end{equation}%
and $\mathbf{X}$ is called a Lie point symmetry, where $\mathbf{X}^{\left[ 1%
\right] }$ is the first extension of the vector field $\mathbf{X}$ in the
jet space \cite{olver,kumei,ibra}.

Therefore for the system (\ref{hm.04})-(\ref{hm.09}) from the symmetry
condition (\ref{hm.15}) we find the Lie point symmetries
\begin{eqnarray*}
X_{1} &=&\partial _{t}~,~X_{2}=\partial _{x}~,X_{3}=t\partial _{x}+\partial
_{u^{1}}~,X_{4}=\partial _{u^{2}}~,~X_{5}=\partial _{u^{3}} \\
X_{6} &=&u^{3}\partial _{u^{2}}-u^{2}\partial _{u^{3}}+B^{3}\partial
_{B^{2}}-B^{2}\partial _{B^{3}}~, \\
X_{7} &=&x\partial _{x}+u^{1}\partial _{u^{1}}+u^{2}\partial
_{u^{2}}+u^{3}\partial _{u^{3}}-2\rho \partial _{\rho }.
\end{eqnarray*}%
For simplicity on our presentation we have omitted the presentation of the
determining equations.

We observe that the admitted Lie point symmetries are the time and space
translation, the Galilean boost in the $x$ direction is described by~$X_{3}$%
, while $X_{4},~X_{5}$ are translation symmetries on the velocity on the
directions $y$ and $z$, the vector field $X_{6}$ is a rotation symmetry and $%
X_{7}$ is a scaling symmetry.

In order to compare the symmetries with that of the MHD system, we assume $%
\xi _{0}=0$ in (\ref{hm.04})-(\ref{hm.09}) from where the symmetry vectors
follows%
\begin{equation*}
X_{1}~,~X_{2}~,~X_{3}~,~X_{4}~,~X_{5}~,~X_{6}~,~X_{7}\text{ and }%
X_{8}=t\partial _{t}+x\partial _{x}\text{.}
\end{equation*}%
Hence, we observe that in the presence of the Hall parameter the scaling
symmetry $X_{8}$ is omitted

The existence of Lie point symmetries for the system (\ref{hm.04})-(\ref%
{hm.09}) it is essential for the determination of similarity solutions and
of conservation laws. In this work we are interested on the determination of
similarity solutions which follow by the application of the Lie invariants
\cite{kumei}. The application of a Lie point symmetry for the reduction of
the system of partial differential equations (\ref{hm.04})-(\ref{hm.09})
lead to a system of ordinary differential equations. In order to determine
all the unique similarity transformations we should calculate Adjoint
representation of the admitted seven-dimensional Lie algebra an after find
the one-dimensional optimal system.

In order to understand this consider the two vector fields
\begin{equation}
\mathbf{Z}=\sum\limits_{A=1}^{n}a_{A}X_{A}~,~\mathbf{W}=\sum%
\limits_{i=1}^{n}b_{A}X_{A}~,~\text{\ }a_{A},~b_{A}\text{ are constants and }%
A=1,2,...,7.  \label{hm.16}
\end{equation}%
The vector fields $Z,~W$ are equivalent if an only
\begin{equation}
\mathbf{W}=Ad\left( \exp \left( \varepsilon _{A}X_{A}\right)
\right) \mathbf{Z~,~}  \label{hm.17}
\end{equation}%
or $W=cZ~,~c=const.~$Operator $Ad\left( \exp \left( \varepsilon
_{A}X_{A}\right) \right) $ is called the Adjoint representation \cite{kumei}%
\begin{equation}
Ad\left( \exp \left( \varepsilon X_{A}\right) \right)
X_{B}=X_{A}-\varepsilon \left[ X_{A},X_{B}\right] +\frac{1}{2}\varepsilon
^{2}\left[ X_{A},\left[ X_{A},X_{B}\right] \right] +...~,  \label{hm.18}
\end{equation}%
where $\left[ X_{A},X_{B}\right] $ is the commutator operator. In Tables \ref%
{tab1} and \ref{tab2} we present the commutators and the Adjoint
representation for the admitted seven-dimensional Lie algebra by the system
1D HMHD system (\ref{hm.04})-(\ref{hm.09}).

\begin{table}[tbp] \centering%
\caption{Commutators of the admitted Lie point symmetries for the 1D HMHD
system}%
\begin{tabular}{cccccccc}
\hline\hline
$\left[ ~,~\right] $ & $\mathbf{X}_{1}$ & $\mathbf{X}_{2}$ & $\mathbf{X}_{3}$
& $\mathbf{X}_{4}$ & $\mathbf{X}_{5}$ & $\mathbf{X}_{6}$ & $\mathbf{X}_{7}$
\\
$\mathbf{X}_{1}$ & $0$ & $0$ & $X_{2}$ & $0$ & $0$ & $0$ & $0$ \\
$\mathbf{X}_{2}$ & $0$ & $0$ & $0$ & $0$ & $0$ & $0$ & $X_{2}$ \\
$\mathbf{X}_{3}$ & $-X_{2}$ & $0$ & $0$ & $0$ & $0$ & $0$ & $X_{3}$ \\
$\mathbf{X}_{4}$ & $0$ & $0$ & $0$ & $0$ & $0$ & $-X_{5}$ & $X_{4}$ \\
$\mathbf{X}_{5}$ & $0$ & $0$ & $0$ & $0$ & $0$ & $X_{4}$ & $X_{5}$ \\
$\mathbf{X}_{6}$ & $0$ & $0$ & $0$ & $X_{5}$ & $-X_{4}$ & $0$ & $0$ \\
$\mathbf{X}_{7}$ & $0$ & $-X_{2}$ & $-X_{3}$ & $-X_{4}$ & $-X_{5}$ & $0$ & $%
0 $ \\ \hline\hline
\end{tabular}%
\label{tab1}%
\end{table}%

\begin{table}[tbp] \centering%
\caption{Adjoint representation for the admitted Lie point symmetries of
the 1D HMHD system}%
\begin{tabular}{cccccccc}
\hline\hline
$Ad\left( \exp \left( \varepsilon X_{A}\right) \right) X_{B}$ & $\mathbf{X}%
_{1}$ & $\mathbf{X}_{2}$ & $\mathbf{X}_{3}$ & $\mathbf{X}_{4}$ & $\mathbf{X}%
_{5}$ & $\mathbf{X}_{6}$ & $\mathbf{X}_{7}$ \\
$\mathbf{X}_{1}$ & $X_{1}$ & $X_{2}$ & $X_{3}-\varepsilon X_{2}$ & $X_{4}$ &
$X_{5}$ & $X_{6}$ & $X_{7}$ \\
$\mathbf{X}_{2}$ & $X_{1}$ & $X_{2}$ & $X_{3}$ & $X_{4}$ & $X_{5}$ & $X_{6}$
& $X_{7}-\varepsilon X_{2}$ \\
$\mathbf{X}_{3}$ & $X_{1}+\varepsilon X_{2}$ & $X_{2}$ & $X_{3}$ & $X_{4}$ &
$X_{5}$ & $X_{6}$ & $X_{7}-\varepsilon X_{3}$ \\
$\mathbf{X}_{4}$ & $X_{1}$ & $X_{2}$ & $X_{3}$ & $X_{4}$ & $X_{5}$ & $%
X_{6}+\varepsilon X_{5}$ & $X_{7}-\varepsilon X_{4}$ \\
$\mathbf{X}_{5}$ & $X_{1}$ & $X_{2}$ & $X_{3}$ & $X_{4}$ & $X_{5}$ & $%
X_{6}-\varepsilon X_{4}$ & $X_{7}-\varepsilon X_{5}$ \\
$\mathbf{X}_{6}$ & $X_{1}$ & $X_{2}$ & $X_{3}$ & $X_{4}\cos \varepsilon
-X_{5}\sin \varepsilon $ & $X_{4}\sin \varepsilon +X_{4}\cos \varepsilon $ &
$X_{6}$ & $X_{7}$ \\
$\mathbf{X}_{7}$ & $X_{1}$ & $e^{\varepsilon }X_{2}$ & $e^{\varepsilon
}X_{3} $ & $e^{\varepsilon }X_{4}$ & $e^{\varepsilon }X_{5}$ & $X_{6}$ & $%
X_{7}$ \\ \hline\hline
\end{tabular}%
\label{tab2}%
\end{table}%

\subsection{One-dimensional optimal system}

The determination of the one-dimensional optimal system is essential in
order to perform a complete classification of the similarity transformations
according to the definition presented in \cite{olver}. As a first step the
invariants $\phi \left( a_{a}\right) ~$of the Adjoint action should be
derived. They are given by the system
\begin{equation}
\Delta _{A}\left( \phi \right) =C_{AB}^{C}a^{B}\frac{\partial }{\partial
a^{A}}\phi \equiv 0  \label{hm.19}
\end{equation}%
where $C_{AB}^{C}$ are the structure constants of the Lie algebra, as they
are presented in Table \ref{tab1}.

Therefore, we end with the system
\begin{equation}
a_{3}\frac{\partial \phi }{\partial a_{2}}=0~,~-a_{1}\frac{\partial \phi }{%
\partial a_{2}}+a_{7}\frac{\partial \phi }{\partial a_{3}}=0~,~-a_{6}\frac{%
\partial \phi }{\partial a_{5}}+a_{7}\frac{\partial \phi }{\partial a_{4}}=0
\end{equation}%
\begin{equation}
-a_{6}\frac{\partial \phi }{\partial a_{4}}+a_{7}\frac{\partial \phi }{%
\partial a_{5}}=0~,~a_{4}\frac{\partial \phi }{\partial a_{5}}-a_{5}\frac{%
\partial \phi }{\partial a_{4}}=0~,
\end{equation}%
\begin{equation}
a_{2}\frac{\partial \phi }{\partial a_{2}}-a_{3}\frac{\partial \phi }{%
\partial a_{3}}-a_{4}\frac{\partial \phi }{\partial a_{4}}-a_{5}\frac{%
\partial \phi }{\partial a_{5}}=0.
\end{equation}%
which gives $\phi =\phi \left( a_{1},a_{6},a_{7}\right) $, that is, the
invariants are the $a_{1},~a_{6},~$ and $a_{7}$.

Consider now the generic symmetry vector%
\begin{equation}
Y=a_{1}X_{1}+a_{2}X_{2}+a_{3}X_{3}+a_{4}X_{4}+a_{5}X_{5}+a_{6}X_{6}+a_{7}X_{7},
\end{equation}%
for the case where $a_{1}a_{6}a_{7}\neq 0$, it follows%
\begin{equation}
\bar{Y}=Ad\left( e^{\varepsilon _{2}X_{2}}\right) Ad\left( e^{\varepsilon
_{3}X_{3}}\right) Ad\left( e^{\varepsilon _{4}X_{4}}\right) Ad\left(
e^{\varepsilon _{5}X_{5}}\right) Y,
\end{equation}%
where for specific values of the free parameters $\varepsilon
_{2},~\varepsilon _{3},~\varepsilon _{4}$ and $\varepsilon _{5}$, the vector
field $\bar{Y}$ takes the form%
\begin{equation}
\bar{Y}=\bar{a}_{1}X_{1}+\bar{a}_{6}X_{6}+\bar{a}_{7}X_{7}.
\end{equation}%
which is the equivalent symmetry vector to the generic field $Y.$

For $a_{1}a_{6}\neq 0$ ,~$a_{7}=0$, with the same approach we find the
invariants $\phi \left( a_{1},a_{3},a_{6}\right) $, from where we get the
equivalent vector field of $Y$ to be%
\begin{equation}
Y^{\prime }=a_{1}^{\prime }X_{1}+a_{3}^{\prime }X_{3}+a_{6}^{\prime }X_{6}.
\end{equation}%
In a similar way we continue and for the rest of the invariants. Therefore,
we find that the one-dimensional optimal system for the 1D HMHD system (\ref%
{hm.04})-(\ref{hm.09}) is consisted by the symmetry vector fields%
\begin{eqnarray*}
&&\left\{ X_{1}\right\} ~,~\left\{ X_{2}\right\} ~,~\left\{ X_{3}\right\}
~,~\left\{ X_{4}\right\} ~,~\left\{ X_{5}\right\} ~,~\left\{ X_{6}\right\}
~,~\left\{ X_{7}\right\} \\
&&\left\{ X_{1}+\alpha X_{2}\right\} ~,~\left\{ X_{1}+\alpha X_{3}\right\}
~,~\left\{ X_{1}+\alpha X_{4}\right\} ~,~\left\{ X_{1}+\alpha X_{5}\right\}
~,~ \\
&&\left\{ X_{1}+\alpha X_{7}\right\} ~,~\left\{ X_{1}+\alpha X_{3}+\beta
X_{4}\right\} ,~\left\{ X_{1}+\alpha X_{3}+\beta X_{5}\right\} ~, \\
&&\left\{ X_{1}+\alpha X_{3}+\beta X_{6}\right\} ~,~\left\{ X_{1}+\alpha
X_{6}+\beta X_{7}\right\} ~~,~\left\{ X_{2}+\alpha X_{4}\right\} ~,~\left\{
X_{2}+\alpha X_{5}\right\} ~, \\
&&\left\{ X_{2}+\alpha X_{6}\right\} ~,~\left\{ X_{2}+\alpha X_{3}+\beta
X_{4}\right\} ~,~\left\{ X_{2}+\alpha X_{3}+\beta X_{5}\right\} ~,\left\{
X_{2}+\alpha X_{3}+\beta X_{6}\right\} ~, \\
&&\left\{ X_{2}+\alpha X_{3}+\beta X_{4}+\gamma X_{5}\right\} ~,~\left\{
X_{2}+\alpha X_{3}+\beta X_{4}+\gamma X_{6}\right\} ~,~\left\{ X_{2}+\alpha
X_{3}+\beta X_{5}+\gamma X_{6}\right\} ~, \\
&&\left\{ X_{3}+\alpha X_{4}\right\} ~,~\left\{ X_{3}+\alpha X_{5}\right\}
~,\left\{ X_{3}+\alpha X_{6}\right\} ~,\left\{ X_{3}+aX_{4}+\beta
X_{5}\right\} ~,~\left\{ X_{3}+aX_{4}+\beta X_{6}\right\} ~, \\
&&\left\{ X_{3}+aX_{5}+\beta X_{6}\right\} ~,~\left\{ X_{4}+\alpha
X_{5}\right\} ~,~\left\{ X_{4}+\alpha X_{6}\right\} ~,~\left\{ X_{5}+\alpha
X_{6}\right\} ~,~\left\{ X_{6}+\alpha X_{7}\right\} \text{.}
\end{eqnarray*}

The Lie invariants which define the similarity transformations for the
one-dimensional optimal system are presented in Tables \ref{tab3}, \ref{tab4}
and \ref{tab5}. In this tables the $\mathcal{R}\left( \theta \right) $
denotes the rotation matrix defined as follows%
\begin{equation*}
\mathcal{R}\left( \theta \right) =%
\begin{pmatrix}
\cos \theta & -\sin \theta \\
\sin \theta & \cos \theta%
\end{pmatrix}%
\text{.}
\end{equation*}

We proceed our analysis with the application of the Lie invariants for the
determination of similarity solutions for the 1D HMHD system.

\begin{table}[tbp] \centering%
\caption{Invariant functions of the one-dimensional optimal system for the 1D HMHD
system (1/3)}%
\begin{tabular}{cc}
\hline\hline
\textbf{Symmetry} & \textbf{Invariants} \\
$X_{1}$ & $x,\mathbf{u,}B^{2},B^{3},\rho $ \\
$X_{2}$ & $t,\mathbf{u,}B^{2},B^{3},\rho $ \\
$X_{3}$ & $t,\frac{x}{t}-u^{1},u^{2},u^{3},B^{2},B^{3},\rho $ \\
$X_{4}$ & $t,x,u^{1},u^{3},B^{2},B^{3},\rho $ \\
$X_{5}$ & $t,x,u^{1},u^{2},B^{2},B^{3},\rho $ \\
$X_{6}$ & $t,x,u^{1}~,~\left( u^{2}\right) +\left( u^{3}\right) ^{2},\left(
B^{2}\right) +\left( B^{3}\right) ^{2},\arctan \left( \frac{u^{3}}{u^{2}}%
\right) -\arctan \left( \frac{B^{3}}{B^{2}}\right) ,~\rho $ \\
$X_{7}$ & $t,\mathbf{u,}B^{2},B^{3},x^{2}\rho $ \\
$X_{1}+\alpha X_{2}$ & $x-\alpha t,\mathbf{u,}B^{2},B^{3},\rho $ \\
$X_{1}+\alpha X_{3}$ & $x-\frac{\alpha }{2}%
t^{2},t-u^{1},u^{2},u^{3},B^{2},B^{3},\rho $ \\
$X_{1}+\alpha X_{4}$ & $x,u^{1},\alpha t-u^{2},u^{3},B^{2},B^{3},\rho $ \\
$X_{1}+\alpha X_{5}$ & $x,u^{1},u^{2},\alpha t-u^{3},B^{2},B^{3},\rho $ \\
\hline\hline
\end{tabular}%
\label{tab3}%
\end{table}%

\begin{table}[tbp] \centering%
\caption{Invariant functions of the one-dimensional optimal system for the 1D HMHD
system (2/3)}%
\begin{tabular}{cc}
\hline\hline
\textbf{Symmetry} & \textbf{Invariants} \\
$X_{1}+\alpha X_{7}$ & $xe^{-\alpha t},e^{-at}\mathbf{u,}%
B^{2},B^{3},e^{-2at}\rho $ \\
$X_{2}+\alpha X_{4}$ & $t,u^{1},ax-u^{2},u^{3},B^{2},B^{3},\rho $ \\
$X_{2}+\alpha X_{5}$ & $t,u^{1},u^{2},ax-u^{3},B^{2},B^{3},\rho $ \\
$X_{2}+\alpha X_{6}$ & $t,u^{1},\mathcal{R}\left( \alpha x\right)
\begin{pmatrix}
u^{2} \\
u^{3}%
\end{pmatrix}%
,\mathcal{R}\left( \alpha x\right)
\begin{pmatrix}
B^{2} \\
B^{3}%
\end{pmatrix}%
,~\rho $ \\
$X_{3}+\alpha X_{4}$ & $t,\frac{x}{t}-u^{1},\alpha \frac{x}{t}%
-u^{2},u^{3},B^{2},B^{3},\rho $ \\
$X_{3}+\alpha X_{5}$ & $t,\frac{x}{t}-u^{1},u^{2},\alpha \frac{x}{t}%
-u^{3},B^{2},B^{3},\rho $ \\
$X_{3}+\alpha X_{6}$ & $t,\frac{x}{t}-u^{1},\mathcal{R}\left( \alpha \frac{x%
}{t}\right)
\begin{pmatrix}
u^{2} \\
u^{3}%
\end{pmatrix}%
,\mathcal{R}\left( \alpha \frac{x}{t}\right)
\begin{pmatrix}
B^{2} \\
B^{3}%
\end{pmatrix}%
,~\rho $ \\
$X_{4}+\alpha X_{5}$ & $t,x,u^{1},u^{2}-\alpha u^{3},B^{2},B^{3},\rho $ \\
$X_{4}+\alpha X_{6}$ & $t,x,u^{1},\left( u^{2}\right) ^{2}+\left(
u^{3}\right) ^{2}+\frac{2}{\alpha }u^{3},~\left( B^{2}\right) +\left(
B^{3}\right) ^{2}~,~\arctan \left( \frac{B^{3}}{B^{2}}\right) -\arctan
\left( \frac{a}{1+au^{3}}\right) ~,~\rho $ \\
$X_{5}+\alpha X_{6}$ & $t,x,u^{1},\left( u^{2}\right) ^{2}+\left(
u^{3}\right) ^{2}+\frac{2}{\alpha }u^{2},~\left( B^{2}\right) +\left(
B^{3}\right) ^{2}~,~\arctan \left( \frac{B^{3}}{B^{2}}\right) -\arctan
\left( \frac{a}{1+au^{2}}\right) ~,~\rho $ \\
$X_{6}+\alpha X_{7}$ & $t,\frac{u^{1}}{x},\frac{i}{2}\left(
u^{2}-iu^{3}\right) x^{-\frac{\alpha +i}{\alpha }},~-\frac{1}{2}\left(
iu^{2}-u^{3}\right) x^{-\frac{\alpha +i}{\alpha }},\mathcal{R}\left( \frac{%
\ln x}{a}\right)
\begin{pmatrix}
B^{2} \\
B^{3}%
\end{pmatrix}%
,~x^{2}\rho $ \\ \hline\hline
\end{tabular}%
\label{tab4}%
\end{table}%

\begin{table}[tbp] \centering%
\caption{Invariant functions of the one-dimensional optimal system for the 1D HMHD
system (3/3)}%
\begin{tabular}{cc}
\hline\hline
\textbf{Symmetry} & \textbf{Invariants} \\
$X_{1}+\alpha X_{3}+\beta X_{4}$ & $x-\frac{\alpha }{2}t^{2},\alpha
t-u^{1},\beta t-u^{2},u^{3},B^{2},B^{3},\rho $ \\
$X_{1}+\alpha X_{3}+\beta X_{5}$ & $x-\frac{\alpha }{2}t^{2},\alpha
t-u^{1},u^{2},\beta t-u^{3},B^{2},B^{3},\rho $ \\
$X_{1}+\alpha X_{3}+\beta X_{6}$ & $x-\frac{\alpha }{2}t^{2},\alpha t-u^{1},%
\mathcal{R}\left( \beta t\right)
\begin{pmatrix}
u^{2} \\
u^{3}%
\end{pmatrix}%
,\mathcal{R}\left( \beta t\right)
\begin{pmatrix}
B^{2} \\
B^{3}%
\end{pmatrix}%
,~\rho $ \\
$X_{1}+\alpha X_{6}+\beta X_{7}$ & $xe^{-\beta t},e^{-\beta
t}u^{1},e^{-\beta t}\mathcal{R}\left( at\right)
\begin{pmatrix}
u^{2} \\
u^{3}%
\end{pmatrix}%
,\mathcal{R}\left( at\right)
\begin{pmatrix}
B^{2} \\
B^{3}%
\end{pmatrix}%
,~e^{2\beta t}\rho $ \\
$X_{2}+\alpha X_{3}+\beta X_{4}$ & $t,\frac{\alpha x}{1+\alpha t}-u^{1},%
\frac{\alpha x}{1+\alpha t}-u^{2},u^{3},B^{2},B^{3},\rho $ \\
$X_{2}+\alpha X_{3}+\beta X_{5}$ & $t,\frac{\alpha x}{1+\alpha t}%
-u^{1},u^{2},\frac{\alpha x}{1+\alpha t}-u^{3},B^{2},B^{3},\rho $ \\
$X_{2}+\alpha X_{3}+\beta X_{6}$ & $t,\frac{\alpha x}{1+\alpha t}-u^{1},%
\mathcal{R}\left( \frac{\beta x}{1+\alpha t}\right)
\begin{pmatrix}
u^{2} \\
u^{3}%
\end{pmatrix}%
,\mathcal{R}\left( \frac{\beta x}{1+\alpha t}\right)
\begin{pmatrix}
B^{2} \\
B^{3}%
\end{pmatrix}%
,~\rho $ \\
$X_{3}+aX_{4}+\beta X_{5}$ & $t,\frac{x}{t}-u^{1},\frac{x}{t}-u^{2},\frac{x}{%
t}-u^{3},B^{2},B^{3},\rho $ \\
$X_{3}+aX_{4}+\beta X_{6}$ & $t,\frac{x}{t}-u^{1},\mathcal{R}\left( \frac{%
\beta x}{t}\right)
\begin{pmatrix}
u^{2} \\
u^{3}%
\end{pmatrix}%
-%
\begin{pmatrix}
\frac{a}{\beta }\cos \frac{\beta x}{t} \\
-\frac{a}{\beta }\sin \frac{\beta x}{t}%
\end{pmatrix}%
,\mathcal{R}\left( \frac{\beta x}{t}\right)
\begin{pmatrix}
B^{2} \\
B^{3}%
\end{pmatrix}%
,~\rho $ \\
$X_{3}+aX_{5}+\beta X_{6}$ & $t,\frac{x}{t}-u^{1},\mathcal{R}\left( \frac{%
\beta x}{t}\right)
\begin{pmatrix}
u^{2} \\
u^{3}%
\end{pmatrix}%
-%
\begin{pmatrix}
\frac{a}{\beta }\sin \frac{\beta x}{t} \\
-\frac{a}{\beta }\cos \frac{\beta x}{t}%
\end{pmatrix}%
,\mathcal{R}\left( \frac{\beta x}{t}\right)
\begin{pmatrix}
B^{2} \\
B^{3}%
\end{pmatrix}%
,~\rho $ \\
$X_{2}+\alpha X_{3}+\beta X_{4}+\gamma X_{5}$ & $t,\frac{\alpha x}{1+\alpha t%
}-u^{1},\frac{\beta x}{1+\alpha t}-u^{2},\frac{\gamma x}{1+\alpha t}%
-u^{3},B^{2},B^{3},\rho $ \\
$X_{2}+\alpha X_{3}+\beta X_{4}+\gamma X_{6}$ & $t,\frac{\alpha x}{1+\alpha t%
}-u^{1},\mathcal{R}\left( \frac{\gamma x}{1+\alpha t}\right)
\begin{pmatrix}
u^{2} \\
u^{3}%
\end{pmatrix}%
-%
\begin{pmatrix}
\frac{\beta }{\gamma }\cos \frac{\gamma x}{1+\alpha t} \\
-\frac{\beta }{\gamma }\sin \frac{\gamma x}{1+\alpha t}%
\end{pmatrix}%
,\mathcal{R}\left( \frac{\gamma x}{1+\alpha t}\right)
\begin{pmatrix}
B^{2} \\
B^{3}%
\end{pmatrix}%
,~\rho $ \\
$X_{2}+\alpha X_{3}+\beta X_{5}+\gamma X_{6}$ & $t,\frac{x}{t}-u^{1},%
\mathcal{R}\left( \frac{\gamma x}{1+\alpha t}\right)
\begin{pmatrix}
u^{2} \\
u^{3}%
\end{pmatrix}%
-%
\begin{pmatrix}
\frac{\beta }{\gamma }\sin \frac{\gamma x}{1+\alpha t} \\
-\frac{\beta }{\gamma }\cos \frac{\gamma x}{1+\alpha t}%
\end{pmatrix}%
,\mathcal{R}\left( \frac{\gamma x}{1+\alpha t}\right)
\begin{pmatrix}
B^{2} \\
B^{3}%
\end{pmatrix}%
,~\rho $ \\ \hline\hline
\end{tabular}%
\label{tab5}%
\end{table}%

\section{Similarity transformations}

\label{sec4}

Before we proceed with the application of similarity transformations for the
determination of exact solutions it is important to mention that the
application of Lie point symmetries for partial differential equations
reducing the number of the independent variables until the reduced system to
be consisted by ordinary differential equations. Thus, not all the Lie
symmetries can play role in the reduction of the system (\ref{hm.04})-(\ref%
{hm.09}) into a system of ordinary differential equations.

In the following we continue by demonstrate with some applications how the
Lie symmetries can be applied for the determination of exact solutions.

\subsection{Symmetry vector $X_{1}-X_{2}$}

\begin{figure}[tbp]
\centering\includegraphics[width=1\textwidth]{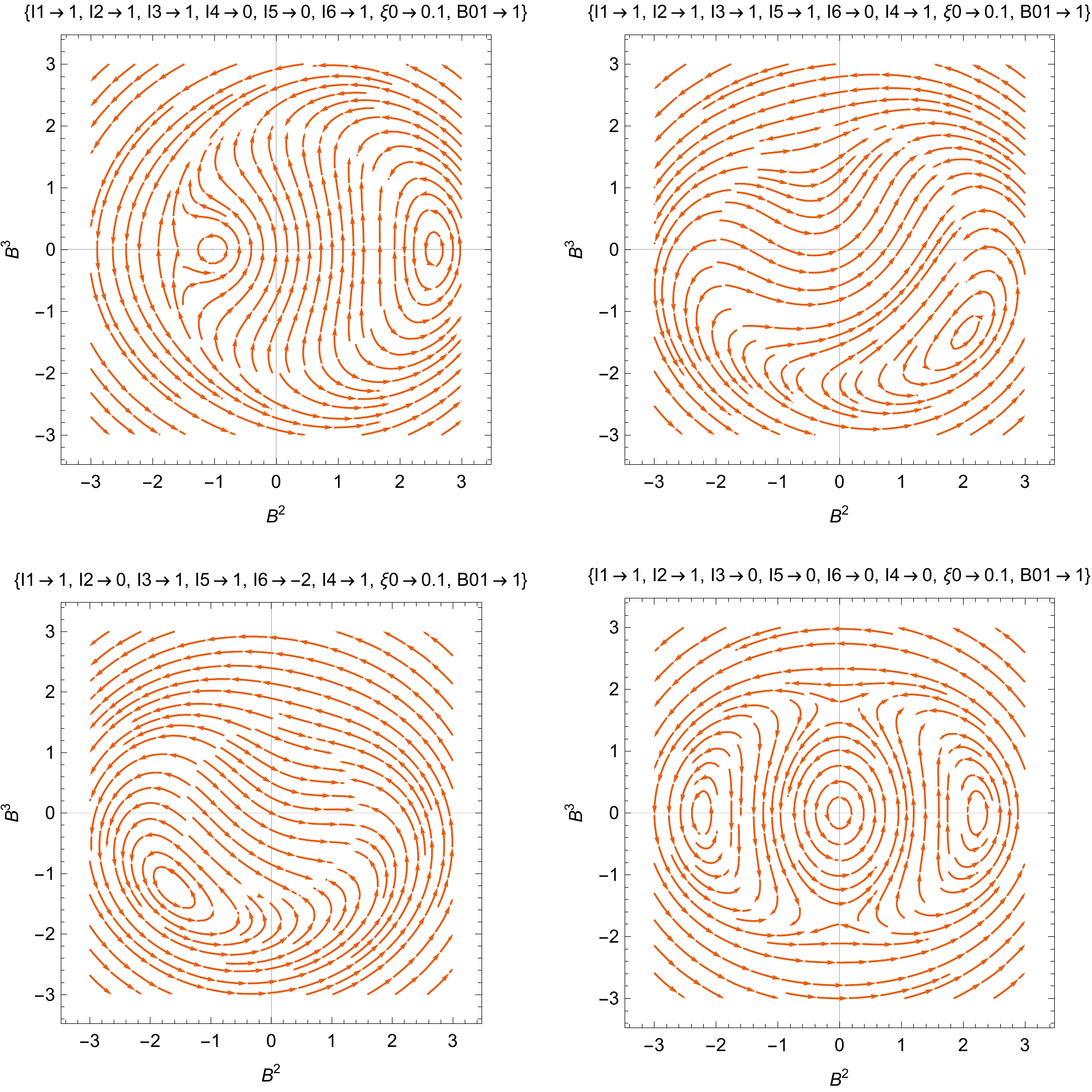}
\caption{Phase space portrait for the system (\protect\ref{hm.30}), (\protect
\ref{hm.32}) for different values of the free parameters. We observe that
there are periodic solutions which indicates the existence of travel-wave
solutions for the system.}
\label{fig1}
\end{figure}

The application of the Lie invariants which follows from the symmetry vector
$\left\{ X_{1}-X_{2}\right\} $ in the 1D HMHD system (\ref{hm.04})-(\ref%
{hm.09}) provides the following system of ordinary differential equations
\begin{eqnarray}
\left( \rho +\rho u^{1}\right) _{,w} &=&0,  \label{hm.20} \\
\left( u^{1}\rho +\rho \left( u^{1}\right) ^{2}+\frac{1}{2}\left( \left(
B^{2}\right) ^{2}+\left( B^{3}\right) ^{2}\right) \right) _{,w} &=&0,
\label{hm.21} \\
\left( \rho u^{2}+\rho u^{1}u^{2}-B^{1}B^{2}\right) _{,w} &=&0,
\label{hm.23} \\
\left( \rho u^{3}\right) _{,t}+\left( \rho u^{3}+\rho
u^{1}u^{3}-B^{1}B^{3}\right) _{,w} &=&0,  \label{hm.24} \\
\left( B^{2}-\left( u^{1}B^{2}-u^{2}B_{0}^{1}\right) +\xi _{0}\frac{B_{0}^{1}%
}{\rho }B_{,x}^{3}\right) _{,w} &=&0,  \label{hm.25} \\
\left( B_{,t}^{3}-\left( u^{3}B_{0}^{1}-u^{1}B^{3}\right) +\xi _{0}\frac{%
B_{0}^{1}}{\rho }B_{,x}^{2}\right) _{,w} &=&0.  \label{hm.26}
\end{eqnarray}%
where $w=t+x$ and the dependent variables are functions of $w$. From
equations\ (\ref{hm.20})-(\ref{hm.24}) it follows%
\begin{eqnarray}
\rho \left( w\right) &=&I_{1}^{2}\left( \left( I_{0}+I_{1}\right) -\frac{1}{2%
}\left( \left( B^{2}\right) ^{2}+\left( B^{3}\right) ^{2}\right) \right)
^{-1}~,  \label{hm.27} \\
u^{1}\left( w\right) &=&\frac{1}{2I_{1}}\left( 2I_{2}-\left( \left(
B^{2}\right) ^{2}+\left( B^{3}\right) ^{2}\right) \right) ~,  \label{hm.28}
\\
u^{2}\left( w\right) &=&\frac{B_{0}^{1}B^{2}+I_{3}}{I_{1}}~,~u^{3}\left(
w\right) =\frac{B_{0}^{1}B^{3}+I_{3}}{I_{1}}~,  \label{hm.29}
\end{eqnarray}%
in which $I_{1},I_{2},I_{3}$ and $I_{4}$ are integration constants. Hence,
for $B^{2}\left( w\right) $ and $B^{3}\left( w\right) $ we find the
first-order ordinary differential equations%
\begin{eqnarray}
B_{,w}^{2} &=&-\frac{I_{1}}{B_{0}^{1}\xi _{0}}B^{3}+\frac{I_{1}\left(
B_{0}^{1}B^{3}+2B_{0}^{1}I_{4}+2I_{5}I_{1}\right) }{B_{0}^{1}\xi _{0}\left(
2\left( I_{1}+I_{2}\right) -\left( \left( B_{2}\right) ^{2}+\left(
B_{2}^{3}\right) \right) \right) }~,  \label{hm.30} \\
B_{,w}^{3} &=&+\frac{I_{1}}{B_{0}^{1}\xi _{0}}B^{2}-\frac{I_{1}\left(
B_{0}^{1}B^{2}+2B_{0}^{1}I_{3}+2I_{6}I_{1}\right) }{B_{0}^{1}\xi _{0}\left(
2\left( I_{1}+I_{2}\right) -\left( \left( B_{2}\right) ^{2}+\left(
B_{2}^{3}\right) \right) \right) }~,  \label{hm.32}
\end{eqnarray}%
with $I_{5},I_{6}$ to be integration constants. The phase portrait of the
dynamical system (\ref{hm.30}), (\ref{hm.32}) is presented in Fig. \ref{fig1}
where we observe the existence of oscillating trajectories \cite{vvs}.

\subsection{Symmetry vector $X_{1}+X_{3}$}

From the point symmetry vector~$X_{1}+X_{3}$ we find
\begin{equation}
I_{1}=\rho u^{1}~,~u^{2}=\frac{B_{0}^{1}B^{2}+I_{2}}{I_{1}}~,~u^{3}=\frac{%
B_{0}^{1}B^{3}+I_{3}}{I_{1}}~,  \label{hm.33}
\end{equation}%
where $\rho =\rho \left( \sigma \right) ,~B^{2}=B^{2}\left( \sigma \right)
~,~B^{3}=B^{3}\left( \sigma \right) ,~$\ $\sigma =x+\frac{t^{2}}{2}$ satisfy
the following system of first-order ordinary differential equations%
\begin{eqnarray}
\rho _{,\sigma } &=&\frac{1}{\xi _{0}B_{0}^{1}I_{1}^{2}}\rho ^{3}\left( \xi
_{0}B_{0}^{1}+I_{5}B^{2}-I_{6}B^{3}\right) ~,~  \label{hm.34} \\
B_{,\sigma }^{2} &=&\frac{1}{\xi _{0}B_{0}^{1}I_{1}}\left( I_{5}I_{1}\rho
+\left( B_{0}^{1}\rho -I_{1}^{2}\right) B^{3}\right) ~,~  \label{hm.35} \\
B_{,\sigma }^{3} &=&-\frac{1}{\xi _{0}B_{0}^{1}I_{1}}\left( I_{6}I_{1}\rho
+\left( B_{0}^{1}\rho -I_{1}^{2}\right) B^{2}\right) ~.  \label{hm.36}
\end{eqnarray}

In analytic solution of the latter system can be written in terms of the
Laurent expansion%
\begin{eqnarray}
\rho \left( \sigma \right) &=&\rho _{0}\sigma ^{-\frac{1}{2}}+\rho
_{1}\sigma ^{-1}+\rho _{2}\sigma ^{-\frac{3}{2}}+...~,  \label{hm.37} \\
B^{2}\left( \sigma \right) &=&B_{0}^{2}\sigma ^{\frac{1}{2}%
}+B_{1}^{2}+B_{2}^{2}\sigma ^{-\frac{1}{2}}+...~,  \label{hm.38} \\
B^{3}\left( \sigma \right) &=&B_{0}^{3}\sigma ^{\frac{1}{2}%
}+B_{1}^{3}+B_{2}^{3}\sigma ^{-\frac{1}{2}}+...~,  \label{hm.39}
\end{eqnarray}%
in which $\rho _{0}=\frac{iI_{1}}{\sqrt{2}}~,~B_{0}^{2}=\frac{i\sqrt{2}I_{1}%
}{B_{0}^{1}\xi _{0}}I_{5},~B_{0}^{3}=\frac{i\sqrt{2}I_{1}}{B_{0}^{1}\xi _{0}}%
I_{6},...$~.

The qualitative evolution of the dynamical variables $\rho ,~B^{2}$ and $%
B^{3}$ are presented for various sets of initial conditions in Fig. \ref%
{fig2}. From the figure we observe a periodic behaviour on the dynamical
variables.

\begin{figure}[tbp]
\centering\includegraphics[width=1\textwidth]{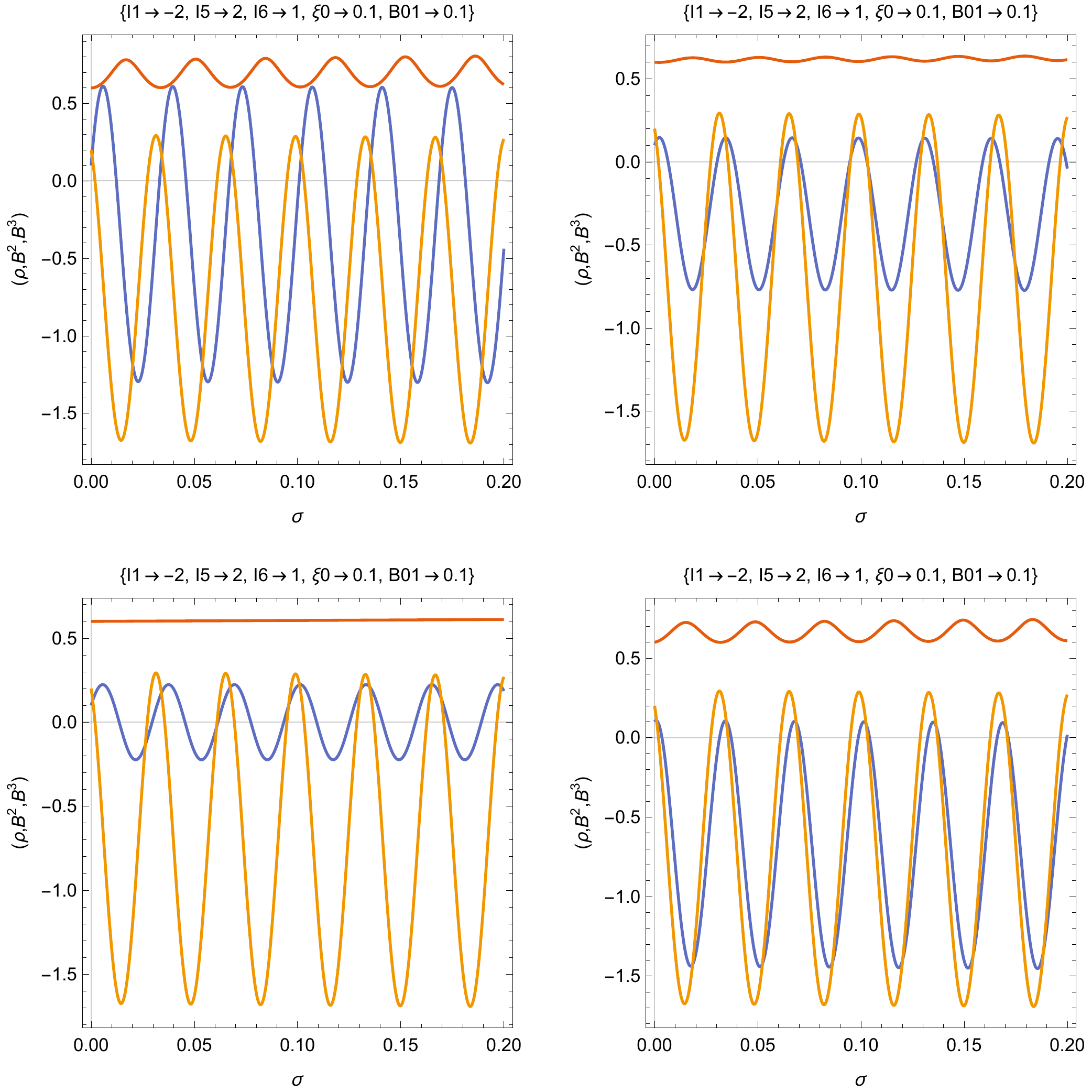}
\caption{Qualitative evolution of the dynamical variables $\protect\rho %
\left( \protect\sigma \right) $ (red line), \ $B^{2}\left( \protect\sigma %
\right) $ (blue line) and $B^{3}\left( \protect\sigma \right) $ (orange
line) for different sets of the free variables. From the plots we observe a
periodic solution, this is a new wave solution.}
\label{fig2}
\end{figure}

\subsection{Symmetry vector $X_{7}$}

Application of the symmetry vector $X_{7}$ in the system (\ref{hm.04})-(\ref%
{hm.09}) provides the closed-form solution%
\begin{eqnarray}
\rho &=&\frac{\rho _{0}t+\rho _{1}}{x^{2}}~,~u^{1}=\frac{\rho _{0}x}{\rho
_{0}t+\rho _{1}}~,~  \label{hm.40} \\
u^{2} &=&\frac{u_{0}^{2}x}{\rho _{0}t+\rho _{1}}~,~u^{3}=\frac{u_{0}^{3}x}{%
\rho _{0}t+\rho _{1}}~,~  \label{hm.42} \\
B^{2} &=&\frac{B_{0}^{1}u_{0}^{2}t+B_{0}^{2}}{\rho _{0}t+\rho _{1}}~,~B^{2}=%
\frac{B_{0}^{1}u_{0}^{3}t+B_{0}^{3}}{\rho _{0}t+\rho _{1}}~.  \label{hm.43}
\end{eqnarray}

\subsection{Symmetry vector $X_{1}+aX_{7}$}

From the vector field $X_{1}+aX_{7}$ we find $z=xe^{-\alpha t},~v=e^{-at}%
\mathbf{u~,~}B^{2},B^{3},\mu =e^{-2at}\rho $, hence by replacing in (\ref%
{hm.04})-(\ref{hm.09}) it follows%
\begin{eqnarray}
I_{1} &=&\rho v^{1}\left( v^{1}-az\right) +\frac{1}{2}\left( \left(
B^{2}\right) +\left( B^{3}\right) ^{2}\right) ^{2}~,  \label{hm.44} \\
I_{2} &=&\rho v^{2}\left( v^{1}-az\right) -B_{0}^{1}B^{2}~,  \label{hm.45} \\
I_{3} &=&\rho v^{3}\left( v^{1}-az\right) -B_{0}^{1}B^{3}~,  \label{hm.46}
\end{eqnarray}%
where $\rho \left( z\right) ,~B^{2}\left( z\right) $ and $B^{3}\left(
z\right) $ satisfy the following system of ordinary differential equation%
\begin{eqnarray}
\left( \rho v^{1}-az\rho \right) _{,z}-az\rho &=&0,  \label{hmm.47} \\
\left( B^{2}v^{1}-B_{0}^{1}v^{2}-\frac{\xi _{0}B_{0}^{1}}{\rho }%
B_{,z}^{2}\right) _{,z}-azB_{,z}^{2} &=&0,  \label{hmm.48} \\
\left( B^{3}v^{1}-B_{0}^{1}v^{3}+\frac{\xi _{0}B_{0}^{1}}{\rho }%
B_{,z}^{3}\right) _{,z}-azB_{,z}^{3} &=&0.  \label{hmm.49}
\end{eqnarray}

\begin{figure}[tbp]
\centering\includegraphics[width=1\textwidth]{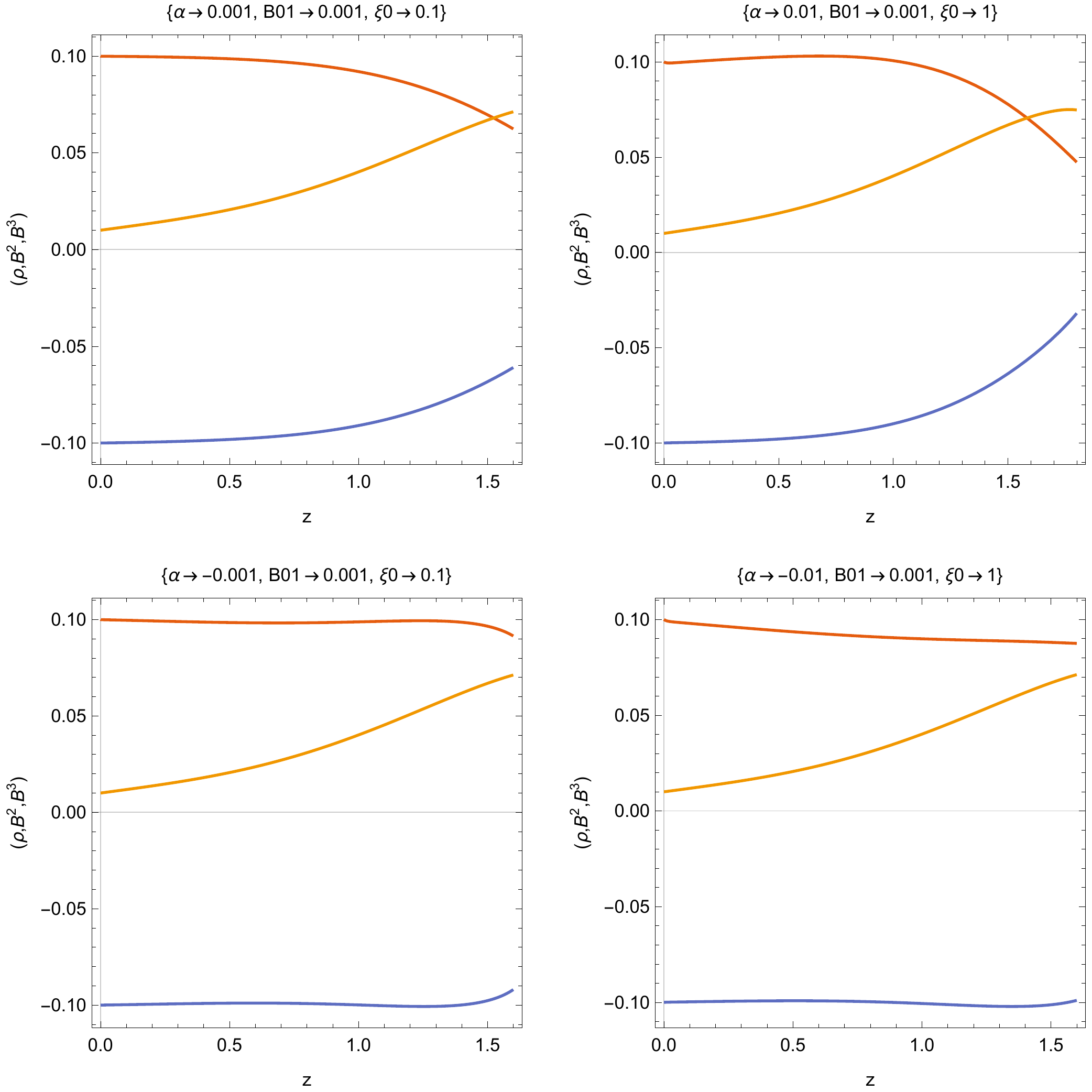}
\caption{Numerical simulation and evolution of the dynamical variables $%
\protect\rho \left( z\right) ~$(red line)$,~B^{2}\left( z\right) $ (blue
line) and $B^{3}\left( z\right) $ (orange line) as it is given by the system
(\protect\ref{hmm.47})-(\protect\ref{hmm.49}). The plots are for different
values of the free parameter. For the initial conditions we considerered $%
\protect\rho \left( 0\right) =0.1$,$~B^{2}\left( 0\right) =0.1$,~$%
B^{3}\left( 0\right) =-0.01$,~$B_{,z}^{2}\left( 0\right) =0.01,$ $%
B_{z}^{3}\left( 0\right) =0.01,~v^{1}\left( 0\right) =0.1,~v^{2}\left(
0\right) =-0.02,~v^{3}\left( 0\right) =0.2$.}
\label{fig3}
\end{figure}

The qualitative evolution of the variables $\rho \left( z\right)
,~B^{2}\left( z\right) $ and $B^{3}\left( z\right) $ as it is given by the
numerical simulation of the system (\ref{hmm.47})-(\ref{hmm.49}) is
presented in Fig. \ref{fig3}.

\subsection{Symmetry vector $X_{3}+aX_{6}$}

From the Lie symmetry vector~$X_{3}+X_{6}$ it follows%
\begin{eqnarray*}
B^{2}\left( t,x\right) &=&-b^{2}\left( t\right) \cos \left( \frac{x}{t}%
\right) +b^{3}\left( t\right) \sin \left( \frac{x}{t}\right) , \\
B^{3}\left( t,x\right) &=&b^{2}\left( t\right) \sin \left( \frac{x}{t}%
\right) +b^{3}\left( t\right) \cos \left( \frac{x}{t}\right) , \\
u^{1}\left( t,x\right) &=&\frac{x}{t}+v^{1}\left( t\right) , \\
u^{2}\left( t,x\right) &=&v^{2}\left( t\right) \sin \left( \frac{x}{t}%
\right) +v^{3}\cos \left( \frac{x}{t}\right) ~, \\
u^{3}\left( t,x\right) &=&v^{2}\left( t\right) \cos \left( \frac{x}{t}%
\right) -v^{3}\sin \left( \frac{x}{t}\right) ~, \\
\rho \left( t,x\right) &=&\rho \left( t\right) ,
\end{eqnarray*}%
while the 1D HMHD system (\ref{hm.04})-(\ref{hm.09}) provides%
\begin{equation}
\rho \left( t\right) =\rho _{0}t^{-1},~v^{1}\left( t\right)
=v_{0}^{1}t^{-1}~,
\end{equation}%
and%
\begin{eqnarray}
ñ _{0}v_{,t}^{2}-ñ _{0}v_{0}^{1}t^{-2}v^{3}-B_{0}^{1}b^{2} &=&0,
\label{hm.50} \\
\rho _{0}v_{,t}^{3}+ñ _{0}v_{0}^{1}t^{-2}v^{2}-B_{0}^{1}b^{3} &=&0,
\label{hm.51} \\
tb_{,t}^{2}-\frac{\xi _{0}B_{0}^{1}}{\rho _{0}}b^{3}+\left(
B_{0}^{1}v^{2}+b^{2}-b^{3}v_{0}^{1}t\right) &=&0,  \label{hm.52} \\
tb_{,t}^{3}+\frac{\xi _{0}B_{0}^{2}}{\rho _{0}}b^{2}+\left(
B_{0}^{1}v^{3}+b^{3}+b^{2}v_{0}^{1}t\right) &=&0.  \label{hm.53}
\end{eqnarray}

Consider the special case where $v_{0}^{1}=0$, then from the two first
equation of the latter system it follows $b^{2}=\frac{ñ _{0}}{B_{0}^{1}%
}v_{,t}^{2}$ and $b^{3}=\frac{ñ _{0}}{B_{0}^{3}}v_{,t}^{3}$. Thus by
replacing in the rest two equations we find%
\begin{eqnarray}
t\frac{ñ _{0}}{B_{0}^{1}}v_{,tt}^{2}-\left( \xi _{0}-\frac{ñ _{0}%
}{B_{0}^{1}}\right) v_{,t}^{2}+B_{0}^{1}v^{2} &=&0~, \\
t\frac{ñ _{0}}{B_{0}^{1}}v_{,tt}^{3}+\left( \xi _{0}+\frac{ñ _{0}%
}{B_{0}^{1}}\right) v_{,t}^{3}+B_{0}^{1}v^{3} &=&0~,
\end{eqnarray}%
with solution which can expressed in terms of the Bessel functions $%
J_{\alpha }\left( t\right) $ and $Y_{\alpha }\left( t\right) $. Note that
for $v_{0}^{1}\neq 0$ system (\ref{hm.50})-(\ref{hm.53}) can be written into
the form of an equivalent system of two linear second-order ordinary
differential equations which means that the system is integrable.

\begin{figure}[tbp]
\centering\includegraphics[width=1\textwidth]{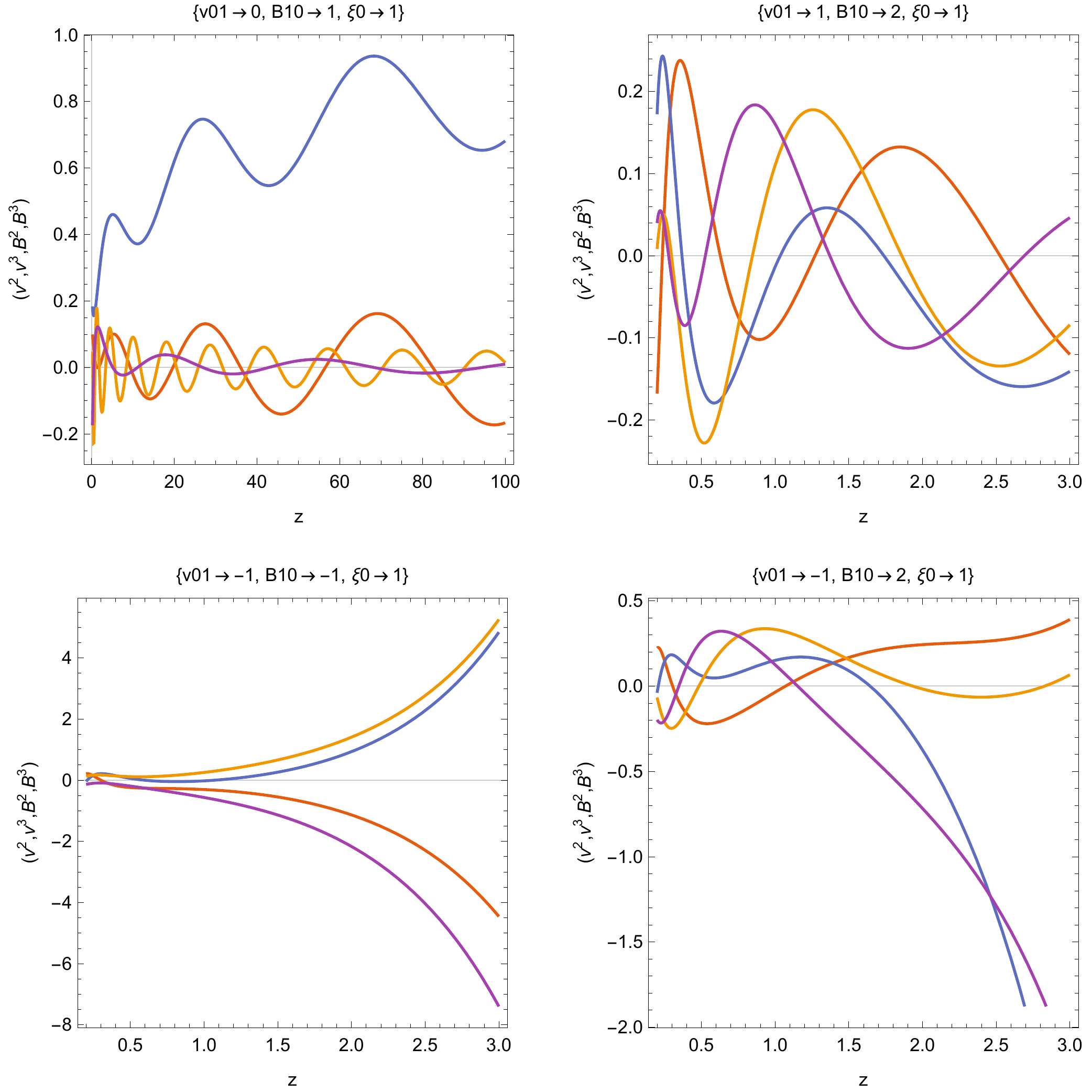}
\caption{Numerical simulation and evolution of the dynamical variables $%
v^{2}\left( t\right) ~$(red line)$,~v^{3}\left( t\right) ~$(blue line) $%
b^{2}\left( z\right) $ (orange line) and $b^{3}\left( z\right) $ (purple
line) as it is given by the system (\protect\ref{hm.50})-(\protect\ref{hm.53}%
). The plots are for different values of the free parameter.}
\label{fig4}
\end{figure}

\begin{figure}[tbp]
\centering\includegraphics[width=1\textwidth]{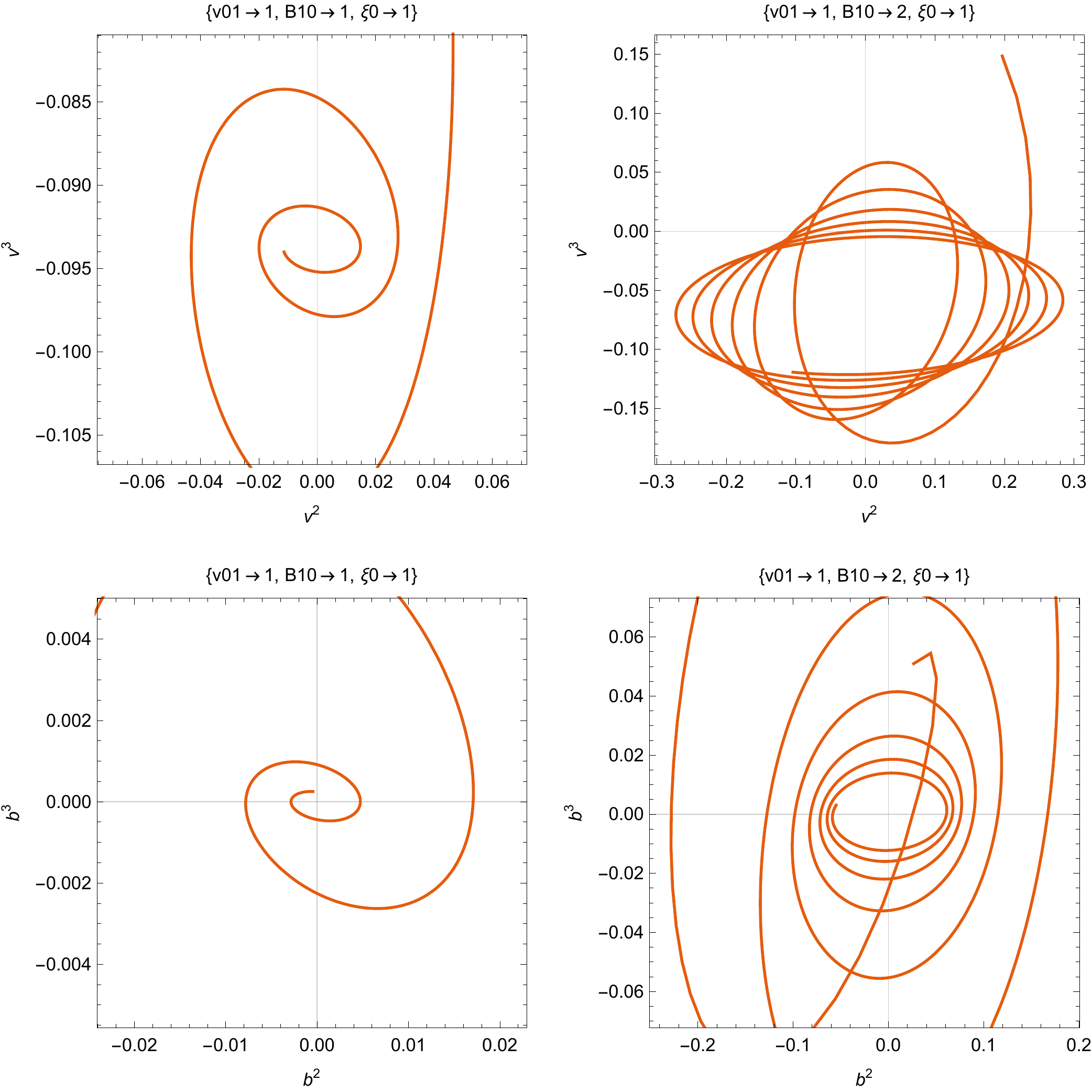}
\caption{Phase space portraits on the variables $\left\{ v^{2}-v^{3}\right\}
$ and $\left\{ b^{2}-b^{3}\right\} $ for the oscillating solutions of Fig.
\protect\ref{fig4}}
\label{fig5}
\end{figure}

In Fig. \ref{fig4} we present the numerical solution of the dynamical system
(\ref{hm.50})-(\ref{hm.53}) while some phase space portraits are given in
Fig. \ref{fig5}. It is clear that the behaviour is of the dynamical
variables has oscillations for $v_{0}^{1}$ and $B_{0}^{1}$ positive.

\section{Conclusions}

\label{sec5}

In this work we studied the group properties the nonlinear one-dimensional
system of MHD included the Hall term. The dynamical system is consisted by
six 1+1 partial differential equations. For this system, we applied the
theory of Sophus Lie and we determined all the possible one-parameter point
transformations in which the HMHD equations are invariant. We found that the
admitted Lie point symmetries form a seven dimensional Lie algebra. The
admitted Lie algebra has common element with that of the MHD system without
the Hall term, thus it admits a smaller number of Lie symmetries for the
equivalent system without the Hall term.

For the admitted Lie symmetries we calculated the commutators and the
Adjoint representations as also the adjoint invariants. These results were
applied in order to determine all the one-dimensional Lie algebras which
consisted the optimal system. We found that the one-dimensional system is
consisted by thirty five independent vector fields. For the latter vector
fields the invariant functions which define the similarity transformations
were determined which are applied for the definition of the similarity
transformations.

Furthermore, we applied the Lie point symmetries to reduce the partial
differential equations into a system of ordinary differential equations and
to study the behaviour of the dynamical variables. Travel-wave and scaling
solutions were found.\

These results contribute to the subject of the application of the Lie point
symmetries to fluid dynamics and specifically on MHD.

\end{document}